\documentclass[epj,nopacs,final]{svjourprep}
\def\preprintnos{\hbox{\phantom{X}}\hbox{NSF-ITP-02-50}\hbox{SACLAY-T02/082}\hbox{TUW-02-14}}
\newcommand\0{\over } \newcommand\2{{1\over2}} 
 \newcommand\6{\partial }

\newcommand{\bea}{\begin{eqnarray}}
\newcommand{\eea}{\end{eqnarray}}
\newcommand{\be}{\begin{equation}}
\newcommand{\ee}{\end{equation}}
\newcommand{\nn}{\nonumber\\ }
\newcommand{\beq}{\begin{eqnarray}}
\newcommand{\eeq}{\end{eqnarray}}
\newcommand{\Tr}{\mathrm{Tr}\,}
\newcommand{\tr}{\mathrm{tr}\,}
\newcommand{\RE}{\,\mathrm{Re}\,}
\newcommand{\IM}{\,\mathrm{Im}\,}
\usepackage{amssymb}
\usepackage{graphicx}
\begin{document}
\title{Comparing different hard-thermal-loop approaches to\\
quark number susceptibilities}
\author{J.-P. Blaizot\inst{1} \and E. Iancu\inst{1,2} \and A. Rebhan\inst{2,3}
}                     
\institute{Service de Physique Th\'eorique, CE Saclay,
        F-91191 Gif-sur-Yvette, France\and 
         Institute for Theoretical Physics, University of California,
         Santa Barbara, CA 93106, U.S.A. \and
        Institut f\"ur Theoretische Physik,
         Technische Universit\"at Wien,
         Wiedner Hauptstr. 8-10,
         A-1040 Vienna, Austria
}
\date{}
\abstract{We compare our previously proposed hard-thermal-loop (HTL) resummed
calculation of quark number susceptibilities
using a self-consistent two-loop approximation to the
quark density with a recent calculation of the same quantity
at the one-loop level in a variant of HTL-screened perturbation theory.
Besides pointing out conceptual problems with the latter approach, we
show that it severely over-includes the leading-order interaction effects
while including none of the plasmon term which after all is the
reason to construct improved resummation schemes.
\PACS{
      {PACS-key}{describing text of that key}   \and
      {PACS-key}{describing text of that key}
     } 
} 
\maketitle
\section{Introduction}
\label{intro}

In view of the ongoing search for quark-gluon plasma signals
in the early stages of
ultrarelativistic heavy-ion collisions, quark number susceptibilities (QNS)
have recently received enhanced attention because of their
direct connection with fluctuations of conserved charges
which could in principle discriminate against a purely
hadronic phase \cite{Asakawa:2000wh,Jeon:2000wg,Prakash:2001xm}.
Concurrently, new results for
QNS have become available from lattice gauge theory
\cite{Gavai,Allton:2002zi}
which considerably improve upon previous
studies \cite{Gottlieb},
and moreover extend them to higher $T/T_c$.
The diagonal\footnote{Offdiagonal QNS are strongly
suppressed in the high-temperature
phase. Still, they give rise to a puzzling discrepancy
between recent analytic and lattice calculations.
The leading-order effect $\propto \alpha_s^3\log(\alpha_s)$
has recently been calculated in Ref.~\cite{Blaizot:2001vr}, 
implying a numerical value of $\sim 10^{-4}$, whereas the
available lattice results are claimed to be consistent with zero
within an accuracy of $\lesssim 10^{-6}$ \cite{Gavai}.
}
QNS are found to increase sharply at the
deconfinement phase transition toward a large percentage of
the ideal gas value $\chi_0=NT^2/3$ for SU($N$) and massless quarks.

Conventionally resummed perturbative results for thermodynamic
quantities, on the other hand, do not seem to be applicable
to account for the observed deviation from ideal-gas behaviour
because of a complete lack of convergence for all temperatures
of interest. In
the case of the free energy, this problem is particularly severe,
because the so-called plasmon contribution $\propto \alpha_s^{3/2}$ is larger
than the leading-order interaction term $\propto \alpha_s$ for all
temperatures $T\lesssim 10^5 T_c$. A similar but less dramatic
problem also occurs with the perturbative result for the
diagonal QNS
\be\label{cc0pt}
{\chi\0\chi_0}=1-2{\alpha_s\0\pi}+8\sqrt{1+{N_f\06}}\left({\alpha_s\0\pi}
\right)^{3/2} + O(\alpha_s^2\log(\alpha_s))
\ee 
for QCD ($N=3$) with $N_f$ quark flavours.
For $N_f=2$ 
the plasmon term overcompensates the term $\propto \alpha_s$ for all
temperatures $T\lesssim 40 T_c$, and only for $T\gtrsim 700 T_c$
does $\chi/\chi_0$ show the
expected growth with temperature, starting from values
extremely close to the ideal-gas result.

In Ref.~\cite{Blaizot:2001vr} we have shown that these problems
can be avoided by a reorganization of perturbation theory which
is based on a self-consistent ($\Phi$-derivable) {\em two-loop}
approximation to the thermodynamic potential \cite{BIR}.
The latter leads to a
nonperturbative expression for entropy and quark density
which can be used to resum
the so-called hard thermal loops (HTL) \cite{Braaten:1990mz,Blaizot:2001nr} 
and particular next-to-leading order corrections thereof.
The results for QNS thus obtained are monotonic functions of
$T/T_c$ which account at least for a sizeable part of the
deviation from the ideal-gas behaviour observed in lattice
calculations for $T/T_c\gtrsim 2T_c$.

Recently, a different approach to resum the effects of HTL
in QNS has been put forward in Ref.~\cite{Chakraborty:2001kx}
which starts from quark number charge correlators. Employing
HTL propagators and vertices at {\em one-loop} order, one finds
 substantially larger deviations from the
ideal-gas limit, seemingly in a
good agreement with the lattice results of Refs.~\cite{Gavai}.

In view of the large efforts invested at present by lattice gauge theorists
to explore effects of small chemical potentials at high temperature
in QCD,
we think it worthwhile to explain the fundamental
differences between our approach and that of 
Ref.~\cite{Chakraborty:2001kx} and why, in our opinion,
the results of the latter are actually misleading.
In particular we show that the one-loop
results of Ref.~\cite{Chakraborty:2001kx}
severely over-include the leading-order interaction effects, while they
contain none of the plasmon effects $\propto \alpha_s^{3/2}$
(which are the source of the problems with conventionally
resummed perturbation theory). Moreover, we point out a
certain technical difficulty that has been overlooked by the authors
of Ref.~\cite{Chakraborty:2001kx}, but has the effect to render
their result ill-defined in a distributional sense.

More importantly even, we comment on a conceptual problem with the
approach followed in Ref.~\cite{Chakraborty:2001kx}, which arises
because the HTL action is no longer used as the effective theory
for soft modes, but is used throughout all of phase space.
Implicitly the definition of the quark number charge operator
is modified such as to no longer conform with the operator
employed in lattice calculations.

Before discussing the approach of Ref.~\cite{Chakraborty:2001kx}
in detail in Sect.~\ref{sec:corr}, we briefly review the QNS
as obtained from HTL-resummed thermodynamic potentials. Sect.~\ref{sec:concl}
summarizes our conclusions.

\section{QNS from resummed thermodynamic potentials}
\label{sec:thpot}

\subsection{Generalities}

The QNS of a given quark flavour is by definition the response of
the quark number density $\cal N$ to an infinitesimal variation of the
associated chemical potential $\mu$,
\be\label{chidef}
\chi={\partial{\cal N}\0\partial\mu}={\partial^2 P\0\partial\mu^2}
=\beta\int d^3x\, \langle \rho(0,{\bf x})\rho(0,{\bf 0}) \rangle
\ee
where $P=(\beta V)^{-1}\log Z$ is the thermodynamic pressure,
$\beta=T^{-1}$ and $\rho=\bar\psi \gamma^0 \psi$.

When thermodynamic consistency is automatic, for example in strict
perturbation theory to a given order in $\alpha_s$, it does not
matter which of the equivalent expressions on the
right-hand side of (\ref{chidef}) is employed. However, when
further resummations are performed that amount to a partial
inclusion of higher-order effects, it does in fact matter.
To set the stage we begin by briefly reviewing the
approaches which focus on
the thermodynamic potential before turning to Ref.~\cite{Chakraborty:2001kx}
which starts from the quark number charge correlator.

Expressed as a functional of full propagators ($D$ for gauge bosons and
$S$ for fermions, and assuming a ghost-free gauge choice) the
thermodynamic potential $\Omega=-PV$ $=-T\log Z$ 
has the form \cite{Luttinger:1960}
\bea\label{LWQCD}
\beta\Omega[D,S]&=&\2 \Tr \log D^{-1}-\2 \Tr \Pi D \nn&&
- \Tr \log S^{-1} + \Tr \Sigma S + \Phi[D,S]
\eea
where $\Phi$ is the sum of 2-particle-irreducible ``skeleton''
diagrams whose lowest-order (2-loop) contributions are

\noindent $\Phi[D,S]={\includegraphics[bb=45 395 400 450,width=4.5cm]{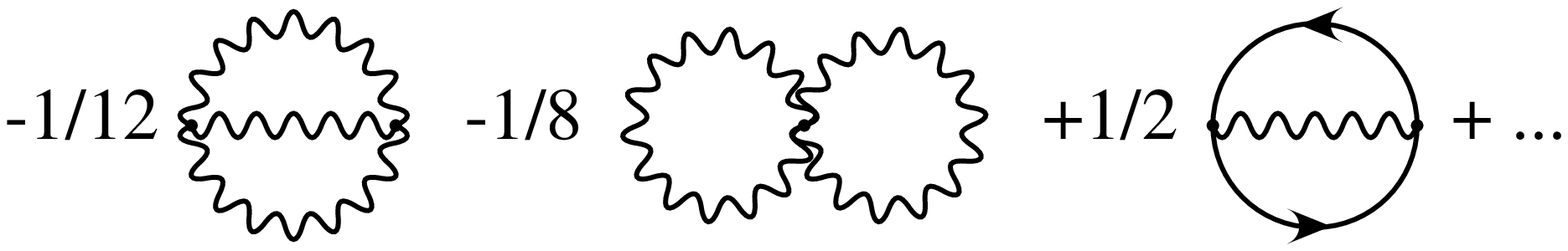}}$

\bigskip\medskip

As a functional of $D$ and $S$, $\Omega$ is subject to the stationarity
condition,
\be\label{staty}
{\delta \Omega[D,S]/\delta D}=0={\delta \Omega[D,S]/\delta S},
\ee
which is equivalent to
\be\label{selfenergies}
{\delta\Phi[D,S]/\delta D}=\2\Pi,\quad
{\delta\Phi[D,S]/\delta S}=\Sigma,
\ee
for the self-energies $\Pi$ and $\Sigma$. Expressing
$\Pi=D^{-1}-D^{-1}_0$ and $\Sigma=S^{-1}-S^{-1}_0$
in terms of bare propagators $D_0$ and $S_0$, the
representation (\ref{LWQCD}) of course reproduces the
ordinary loop expansion.

For example, the leading-order interaction terms $\propto \alpha_s$
are given by the 2-loop diagrams in $\Phi$, whereas single
powers of the self-energy insertions in a propagator cancel
out in the first four terms of the right-hand side of eq. (\ref{LWQCD}).

Ordinary perturbation theory, however, has infrared problems
at finite temperature if the repeated self-energy insertions contained in the
term $\2\Tr\log D^{-1}$ are expanded out perturbatively.
These can be remedied by a resummation
of the leading order Debye mass
\be\label{MD}
\hat m^2_D=(2N+N_f)\,\frac{g^2T^2}{6}+\sum_i{g^2\mu_i^2\02\pi^2}
\ee
in the (chromo-)electrostatic propagator, where $g^2=4\pi\alpha_s$
(though new infrared problems arise at order $\alpha_s^3$).
Expanded in powers of $g$, the resummation of the Debye mass
in $\2\Tr\log D^{-1}$ gives rise to the so-called plasmon
term in the pressure
\be\label{P3}
P_3=N_g T m_D^3/(12\pi).
\ee
It is this term which is responsible for the dramatic deterioration of the
apparent convergence of a perturbative expansion of $P$ in $g$
at finite temperature, and, as remarked in the introduction,
to a somewhat lesser degree for QNS which can be derived from the
pressure.

\subsection{Screened (HTL) perturbation theory}
 
The loss of apparent convergence upon inclusion of the plasmon
term in the pressure is in fact generic and also occurs in a
simple scalar $\varphi^4$ theory \cite{Drummond:1997cw}. This problem
arises
as soon as finite-temperature contributions are expanded out
in powers of the coupling, which is necessary for the
standard ultraviolet renormalization programme to become applicable.
In order to avoid this, it has been proposed \cite{Karsch:1997gj} to reorganize
perturbation theory by adding screening masses to the classical Lagrangian
and to subtract them as counter-terms, but in contrast to
the usual resummation programme at finite temperature 
\cite{Braaten:1990mz,Blaizot:2001nr},
this is done for both hard and soft momentum regimes.
This in fact alters the ultraviolet structure of the theory, but when
combined with
a simple minimal subtraction of the additional divergences this
resummation appears to significantly 
improve the apparent convergence of thermal
perturbation theory.

In Refs.~\cite{ABS,Baier:1999db}
this approach has been extended to QCD at one-loop level. It amounts
to keeping only the logarithmic terms in (\ref{LWQCD}) and replacing
$D$ and $S$ by the HTL propagators,
\be\label{oneloopHTLpt}
\beta\Omega^{1-loop-HTL}=\2 \Tr \log \hat D^{-1}- \Tr \log \hat S^{-1}
\ee
where hatted quantities refer to HTL.
If the thermal mass parameter
in the HTL are exactly the lowest-order ones, this includes
the correct plasmon term (\ref{P3}) without causing the pressure
to exceed the ideal-gas value. However, the leading-order interaction
pressure $\propto \alpha_s$ is over-included by a factor of 2.\footnote{As
explained in the last paper of Ref.~\cite{BIR},
Ref.~\cite{ABS}
had an even stronger over-inclusion due to an inconsistent
use of dimensional regularization.}

In order to have both, the leading-order term $\propto \alpha_s$
and the plasmon term $\alpha_s^{3/2}$ included correctly, it is
necessary to go to two-loop order. 
Starting from two-loop order,
one can turn this so-called ``HTL perturbation theory''
into a variational perturbation theory, where the HTL action
is no longer used as an effective theory for soft modes as in standard
HTL resummation \cite{Braaten:1990mz,Blaizot:2001nr}, but just as a gauge
invariant mass term which is then eliminated by a principle
of minimal sensitivity.

Because of the HTL action involves non-local self-ener\-gies and
vertices, this optimization of perturbation theory is extremely
difficult and has only recently been carried through for QCD
to 2-loop order
\cite{Andersen:2002ey}. The results are a clear improvement
over conventionally resummed perturbation theory as the resummed
pressure remains below the ideal-gas limit despite a full
inclusion of the contributions through order $\alpha^2\log(\alpha_s)$.
However they appear to account for less than half of the
deviation from ideal-gas behaviour that is observed on the lattice.

The main virtues of this approach are its systematic
nature and manifest gauge invariance. From a physical point of view,
a possible weakness is that
the HTL action is used uniformly for soft and hard momenta, whereas
the HTL action is accurate only for soft momenta, and for hard ones
only in the vicinity of the light-cone. 
A related problem is that the artificial UV divergences that are
introduced 
involve new subtraction
scheme dependences.
While these start to be suppressed by powers of $\alpha_s$
only at the (rather forbidding)
three-loop order \cite{Rebhan:2000uc}, these additional scheme dependences
turn out to be numerically rather weak in the two-loop result for QCD.

\subsection{HTL-resummation of the 2-loop $\Phi$-derivable
entropy and density}

While HTL-screened perturbation is in principle rather generally
applicable, we have found,
following up an observation made in Ref.~\cite{Vanderheyden:1998ph},
that specifically for the first
derivatives of the thermodynamic potential
one can derive remarkably simple expressions from 
a self-consistent 2-loop
approximation to the skeleton expansion (\ref{LWQCD})
of the QCD thermodynamic potential. Because of the
stationarity property, these derivatives act only on the
explicit statistical distribution functions, and not also on those
contained in propagators and self-energies. Moreover, after differentiation,
the contribution from $\Phi^{2-loop}$ just cancels part of the
second and fourth term on the right-hand-side 
of (\ref{LWQCD}). The derivatives with
respect to temperature and chemical potential give entropy
and quark densities, respectively, reading
\cite{BIR}
\bea
\label{S2loop}
{\cal S}&=&-\tr \int{d^4k\0(2\pi)^4}{\6n\0\6T} \left[ \IM 
\log D^{-1}-\IM \Pi\RE D \right] \nn
&&-2\,\tr \int{d^4k\0(2\pi)^4}{\6f\0\6T} \Bigl[ \IM
\log S^{-1}-\IM \Sigma \RE S \Bigr], \\
\label{N2loop}
{\cal N}&=&-2\,\tr \int{d^4k\0(2\pi)^4}{\6f\0\6\mu} \left[ \IM
\log S^{-1}-\IM \Sigma \RE S \right],\qquad
\eea
where $D$ and $S$ are determined by one-loop gap equations
(\ref{selfenergies}) obtained by restricting $\Phi$ to two-loop order.

Although these equations have the form of one-loop expressions
involving dressed propagators, they include all the two-loop
contributions, but incorporated in the spectral properties
of the quasi-particles described by the dressed propagators.
This implies that both the leading-order
interaction terms $\propto \alpha_s$ and the plasmon effect
$\propto \alpha_s^{3/2}$ are completely taken into account
as soon as $D$ and $S$ are evaluated
to sufficient accuracy.

Because the expressions (\ref{S2loop}) and (\ref{N2loop}) are
manifestly ultraviolet finite, they can be used to resum
the effects of HTL without the
necessity of subsequent expansions and truncations.\footnote{For
a different approach based on the pressure see also 
Ref.~\cite{Peshier:2000hx}.} At soft
momenta, the HTL are valid expressions to the actual full
propagators that one would have to use in a self-consistent
scheme; at hard momenta it turns out that to leading order
the above expressions only probe the vicinity of the light-cone
where HTL self-energies remain accurate. The next-to-leading
order effect, which is the plasmon effect, turns out to be
to one part covered by HTL resummation in the soft 
regime,
and to the remaining part 
by corrections to the so-called asymptotic thermal
masses from HTL-resummed one-loop diagrams.

\begin{figure}[t]
\centerline{\includegraphics[bb=180 425 285 475,width=2.5cm]{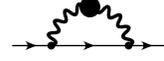}}
\caption{Next-to-leading order corrections to the asymptotic
fermion mass}
\label{Mascorr}
\end{figure}

In the case of the quark number density functional, from which the QNS
can be derived, it turns out that all of the plasmon effect is
associated with next-to-leading order corrections to the asymptotic
fermion mass, shown in Fig.~\ref{Mascorr}.

If only the HTL approximation to the fermion propagator is employed,
the quark number density still 
contains the complete leading-order interaction effects $\propto\alpha_s$,
and -- since it need not be expanded out perturbatively --
also subsets of higher-order effects, namely those associated
with repeated HTL insertions.

In Ref.~\cite{Blaizot:2001vr} we have evaluated the QNS obtained
by taking the derivative of (\ref{N2loop}) with respect to $\mu$,
both in the HTL approximation and in a next-to-leading approximation
which incorporates the plasmon effect through the corrections to
the asymptotic fermion mass from the diagram of Fig.~\ref{Mascorr}.

\section{QNS from HTL-resummed charge correlators}
\label{sec:corr}

In Ref.~\cite{Chakraborty:2001kx} an HTL-resummed QNS has been
constructed by starting from the charge correlator
\be\label{chidef2}
\chi=\beta\int d^3x\, \langle \rho(0,{\bf x})\rho(0,{\bf 0}) \rangle
\equiv \beta\int d^3x\, \Pi_{00}^>(0,{\bf x})
\ee
where $\Pi_{\mu\nu}^>$ is the current-current correlator of
a given flavour charge (suitable linear combinations of such
quantities give the correlators of electric charge and baryon number).

In Fourier-space one has \cite{McLerran:1987pz,Kunihiro:1991qu,Hatsuda:1994pi}
\be\label{chipi00}
\chi=\lim_{k\to 0}\beta \int_{-\infty}^\infty {d\omega\02\pi}
\Pi_{00}^>(\omega,k)
\ee
where the
well-known fluctuation-dissipation theorem 
allows one to write (in the notation of \cite{Blaizot:2001nr})
\be
\Pi_{\mu\nu}^>(\omega,k) = 
-{2\01-e^{-\beta\omega}} \IM \Pi_{\mu\nu}^R(\omega,k)
\ee
with $\Pi_{\mu\nu}^R$  the retarded response function.

\begin{figure}
\centerline{\includegraphics[width=6cm]{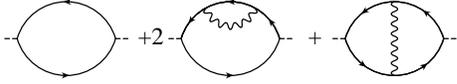}}
\caption{Lowest-order contributions to (\ref{chipi00}) in
bare perturbation theory.}
\label{ccbare}
\end{figure}

In conventional perturbation theory the first few diagrams contained
in (\ref{chipi00}) are shown in Fig.~\ref{ccbare}.\footnote{%
One-particle-reducible diagrams, which in principle contribute,
vanish because of the tracelessness of colour matrices.} The
first diagram gives the ideal gas value, and the leading-order
interaction term $\propto \alpha_s$ in (\ref{cc0pt}) is given by the two-loop
order diagrams. The plasmon effect $\propto \alpha_s^{3/2}$ comes
from those higher-order diagrams which correspond to repeated
self-energy insertions into the gluon lines of the two-loop diagrams
of Fig.~\ref{ccbare}.

Calculating $\chi$ from (\ref{chipi00}) is in fact a bit more
involved than starting from the thermodynamic potential as
also noticed in Ref.~\cite{McLerran:1987pz}. Because charge
conservation implies $\omega\Pi_{00}^>(\omega,0)=0$ one has
$\lim_{k\to0}\Pi_{00}^>(\omega,k) \sim \delta(\omega)$.

If, for example following HTL-screened perturbation theory,
one is interested in the one-loop contribution
arising from dressing the propagators in Fig.~\ref{ccbare},
this will spoil this behaviour. In order to have charge
conservation one needs to employ HTL vertices in addition
to HTL propagators, which is precisely what the authors of
Ref.~\cite{Chakraborty:2001kx} have proposed.

However, this raises an important conceptional problem:
in effect this use of the HTL vertices 
replaces the ordinary charge operator $\bar\psi\gamma^0\psi$
by the non-local object $\bar\psi(\gamma^0+\hat\Gamma^0)\psi$
derivable from the non-local HTL action.
The correspondingly re-defined QNS is therefore
no longer directly related to the quantity defined in (\ref{chidef2})
and measured in lattice simulations.

This is in fact a problem only because the HTL action is no
longer used as an effective theory appropriate for soft
momentum scales, but is used equally for soft and hard momenta.
As an effective theory, obtained after integrating out the
hard momenta and used for soft modes only, 
the appropriate charge operator is indeed
the non-local quantity involving the HTL vertex $\hat\Gamma^0$.
It is this operator which enters in a perturbative matching
to the full theory. But replacing the ordinary
charge operator by the HTL-dressed one for all momenta clearly
corresponds to abandoning the definition (\ref{chidef2}) for the QNS.

\begin{figure}
\centerline{\includegraphics[width=6.5cm]{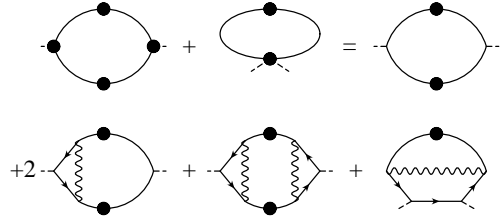}}
\caption{One-loop contributions to (\ref{chipi00}) in
HTL-screened perturbation theory. The vertex parts built
from bare propagators are understood to be evaluated in
the HTL approximation.}
\label{ccHTL1}
\end{figure}

Leaving this issue aside for now, we continue by analy\-sing the
diagrammatic content of $\chi$ in HTL-screened perturbation theory.
Since HTL vertices are strictly one-loop quantities, 
to one-loop order it is as shown in Fig.~\ref{ccHTL1}.

\begin{figure}
\centerline{\includegraphics[width=6.5cm]{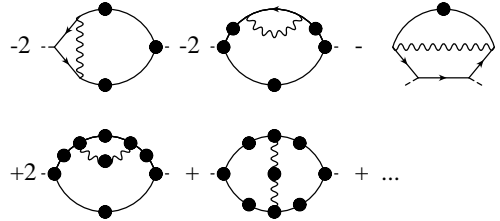}}
\caption{Two-loop contributions to (\ref{chipi00}) in
HTL-screened perturbation theory which contribute to the
leading-order interaction terms $\propto \alpha_s$.}
\label{ccHTL2}
\end{figure}

However, while all the topologies that are present in the
two-loop diagrams of Fig.~\ref{ccbare} also appear in
the diagrams of Fig.~\ref{ccHTL1}, their combinatorial
factors are different. 

This shows that in the one-loop HTL approximation
to (\ref{chipi00}) the $\alpha_s$-contributions are all
overcounted. The second diagram of Fig.~\ref{ccbare} is
contained with correct combinatorics
in the first diagram of the right-hand side of
Fig.~\ref{ccHTL1} but appears another time through
the HTL 4-vertex; the third diagram of Fig.~\ref{ccbare} is seen
to be over-included by a factor of 2. Moreover, because the HTL
approximation for the undressed
self-energies and vertex subdiagrams in Fig.~\ref{ccHTL1}
does not provide the complete
leading-order terms for hard inflowing momenta, this leads
to a further source of incompleteness of the terms $\propto \alpha_s$.

The authors of Ref.~\cite{Chakraborty:2001kx} do not specify how to extend
their approach to two-loop order. It is clear, however, that
the correct counting is restored
only when the modification of the quark number charge is
undone by HTL counter-terms and all two-loop diagrams are added.
The relevant diagrams for completing the
order $\alpha_s$ result are shown in Fig.~\ref{ccHTL2},
where the first and the third diagram correspond to HTL counter-terms
to the charge operator.\footnote{The 2nd and 4th diagram
have opposite combinatorial factors but may also contribute
to order $\alpha_s$ because of the incompleteness of the
HTL approximation for hard loop momenta.}
This means that the
definition of the charge operator in (\ref{chidef2}) has to be
modified order by order to approach to standard definition of QNS
at least at infinite loop order.

This also shows that the plasmon term
$\propto \alpha_s^{3/2}$, which 
is the reason for seeking improvements of conventionally resummed perturbation
theory, only appears in the two-loop order diagrams
of HTL-scree\-ned perturbation theory, namely through the dressed
vector boson lines with a blob. The vector boson lines
within the HTL vertices of Fig.~\ref{ccbare} 
are not dressed and thus do not capture anything of the plasmon effect.

\begin{figure}[t]
\centerline{  
\includegraphics[bb=0 210 540 540,width=0.42\textwidth]{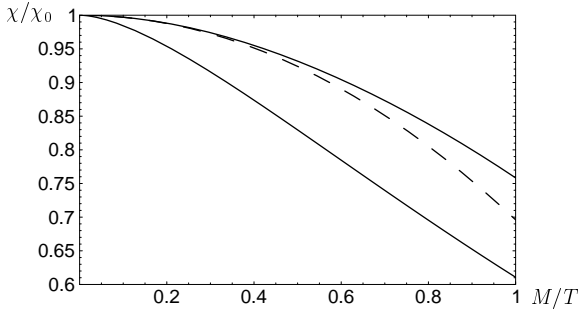}
}
\caption{$\chi/\chi_0$ as a function of $M/T$. The dashed line is
the perturbative result to order $g^2$, the upper full line is
the 2-loop $\Phi$-derivable approximation evaluated with HTL
propagators, the lower full line is the 1-loop HTL result
of Ref.~\cite{Chakraborty:2001kx}.}
\label{fig:cc0}
\end{figure}

We are now in a position to compare with the HTL-resummation
approaches discussed in the previous section.

In one-loop HTL-screened perturbation theory along the lines
of Refs. \cite{ABS,Baier:1999db}
the leading-order interaction term to the thermodynamic potential
is over-included by a factor
2, but the plasmon effect is complete (as long as only the
leading-order HTL mass parameter is used; including higher-order
corrections in the latter would spoil this).
The QNS have not been calculated in this approach, but the same
pattern would apply, provided the perturbative HTL masses
are inserted before differentiating with respect to $\mu$.\footnote{Giving
up thermodynamic consistency, one could think of identifying
the mass parameter in HTL-screened perturbation theory only
after this differentiation. This would in fact give a correct
leading-order interaction term, but would loose the plasmon effect
(and not completely reproduce the terms of order $\alpha_s^2$ and
higher contained in (\ref{N2loop})).}

On the other hand, in the 2-loop $\Phi$-derivable quark density
(\ref{N2loop}), the leading-order interaction term is already
correctly included when evaluating it in the HTL approximation
but the plasmon term is absent
(it arises exclusively from next-to-leading order corrections
to the asymptotic (hard) thermal fermion mass).

In Fig.~\ref{fig:cc0}, the QNS we have obtained from (\ref{N2loop}) 
in the HTL approximation \cite{Blaizot:2001vr} is evaluated as
a function of the fermionic plasmon mass\footnote{For plots
in terms of $\alpha_s$ or $T/T_c$ see Ref.~\cite{Blaizot:2001vr}}
$M/T$ and compared with a numerical evaluation of the one-loop
HTL result reported in Ref.~\cite{Chakraborty:2001kx}.
While our result shows slightly slower deviation from the ideal-gas limit
than the strictly perturbative result to order $\alpha_s^1$,
the result of Ref.~\cite{Chakraborty:2001kx} has considerably stronger
deviations because of the over-inclusion of the
leading-order interaction term.

\begin{figure}[t]
\centerline{\includegraphics[bb=70 230 540 750,width=0.42\textwidth]{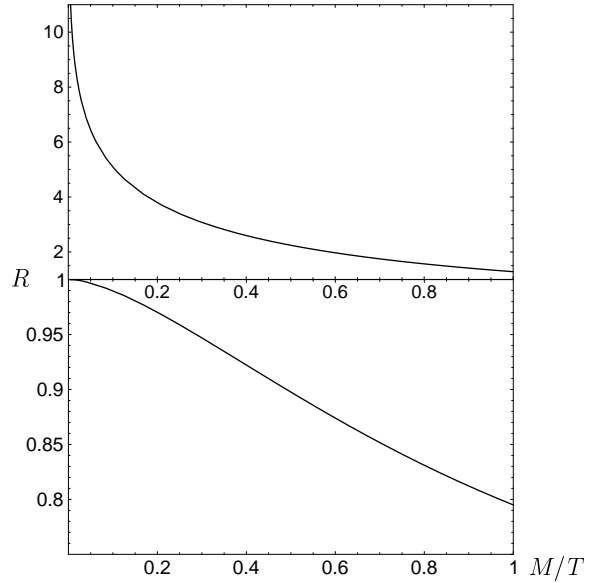}}
\caption{$R=(\chi-\chi_0)/(\chi^{(2)}_{p.th.}-\chi_0)$ as a function of $M/T$
for the 1-loop HTL result
of Ref.~\cite{Chakraborty:2001kx} (upper half of the plot) and
for the 2-loop $\Phi$-derivable approximation evaluated with HTL
propagators (lower half of the plot -- note the different scale!). 
The perturbative result to order $g^2$
corresponds to the value 1. Only the 2-loop HTL result approaches
1 in the limit $M/T\to0$, where eventually perturbation theory
should be reproduced; the 1-loop HTL result of Ref.~\cite{Chakraborty:2001kx}
is seen to over-include the leading-order interaction effect
by a factor which diverges logarithmically as $M/T\propto g\to0$.}
\label{fig:c2}
\end{figure}

In Fig.~\ref{fig:c2} we consider the expression
\be\label{R}
R \equiv (\chi-\chi_0)/(\chi^{(2)}_{p.th.}-\chi_0)
\ee
which measures the deviation of the interaction part of $\chi$ from
the perturbative result $\chi^{(2)}_{p.th.}$ 
to order $\alpha_s$,
and plot the respective results, again as a function of $M/T$.
While our result obtained from (\ref{N2loop}) 
in the HTL approximation \cite{Blaizot:2001vr} goes to 1 in
the limit of a weakly coupled theory, the effective over-inclusion
of the leading-order interaction term of Ref.~\cite{Chakraborty:2001kx}
{\em diverges} in this limit. In fact, it turns out that
the result reported in Ref.~\cite{Chakraborty:2001kx}
involves a 
contribution $\propto (M/T)^2\log(M/T) \sim
\alpha_s \log(\alpha_s)$, which
does not exist in the correct perturbative expansion.
This over-inclusion problem is therefore much more severe
than in the case where 1-loop HTL-screened perturbation theory 
is applied to the thermodynamic potential \cite{ABS,Baier:1999db}.

This clearly shows that the one-loop HTL resummation of
the charge-charge correlator cannot be compared with
either our results, which are based on the two-loop expression
(\ref{N2loop}), or perturbation theory which it seeks to
improve upon. In our opinion it is also completely premature
to compare with the available lattice results on QNS, because
the two-loop contributions of Fig.~\ref{ccHTL2} will
have to correct for the enormous over-inclusion of terms $\propto \alpha_s$.

But it appears to be questionable whether a two-loop HTL-screened
perturbation theory calculation of (\ref{chipi00}) is at all practicable.
There are in fact certain technical problems with the result
reported in Ref.~\cite{Chakraborty:2001kx} already at one-loop order. 
In eq.~(34) of \cite{Chakraborty:2001kx} one can see that the result
for $\Pi_{00}^>(\omega,{\bf 0})$ is proportional to the integral
\bea\label{sick}
&&\int d^3k \int dx \int dx' n_F(x) n_F(x') \rho_+(x,k)\rho_-(x,k)\nn
&&\times {(\omega-x-x')^2\delta(\omega-x-x')\0\omega^2}
\eea
where $\rho_\pm(x,k)$ are the HTL spectral functions for
the two fermionic quasi-particle branches of the HTL approximation.
In Ref.~\cite{Chakraborty:2001kx} 
the latter are used to put $x=-x'$, so that the second line
of (\ref{sick}) is reduced to $\delta(\omega)$ in conformity
with the expectations from charge conservation.
However, this term is clearly ill-defined and might with equal
justification be put to zero identically as is suggested by the way
we have written it. In order to have a well-defined expression,
it seems to be necessary to keep the external spatial momentum
different from zero and take the limit to zero only after
having performed the integral over $\omega$ (as demanded
by (\ref{chipi00})).
But that would make its evaluation in HTL perturbation theory
a hopelessly difficult task, already at one-loop order.

\section{Conclusions}
\label{sec:concl}

In this paper, we have discussed various possibilities for
HTL-resummation in the calculation of QNS and have in particular
analysed the recent proposal of Ref.~\cite{Chakraborty:2001kx}.
We have shown that the resummed one-loop calculation presented there
severely over-includes the leading-order interaction terms, while
not including anything of the plasmon effect, both of which would
be corrected only at two-loop order. 
Thus only the latter should be viewed as an improvement over
ordinary perturbation theory and used as such in a comparison
with the available lattice results.

However, because of the technical problems mentioned at the end of the
previous section, it would seem to be more sensible to calculate
the QNS through a 2-loop HTL-screened perturbation theory evaluation
of the pressure along the lines of Ref.~\cite{Andersen:2002ey}.

On the other hand, the HTL-resummed calculation of QNS of 
Ref.~\cite{Blaizot:2001vr} is based on a 2-loop
$\Phi$--derivable approximation and does include correctly 
both the leading-order interaction effect, $\sim \alpha_s$, 
and the next-to-leading-order one, $\sim \alpha_s^{3/2}$ (together with
an infinite series of higher order effects due to HTL). 
The results in Ref.~\cite{Blaizot:2001vr} show the same
trend as the lattice results, 
but a significant difference still remains,
which calls for further studies, both on the analytic side,
by further improving the resummation schemes, and on the lattice
side, by increasing the reliability of the numerical results.

\section*{Acknowledgements}
J.-P.~Blaizot acknowledges fruitful discussions with R. Ga\-vai and S. Gupta
during a stay at the TIFR supported by the project IFPAR 2104.
E.~Iancu and A.~Rebhan would like to thank the organizers of the program 
{\it ``QCD and Gauge 
Theory Dynamics in the RHIC Era''} at the ITP of the UCSB,
and its staff for hospitality. This research was supported in 
part by the NFS under Grant No. PHY99-07949, and by the 
FWF under Grant No. P14632-TPH.

\end{document}